\documentclass[twocolumn,superscriptaddress,showpacs]{revtex4-1}
\usepackage{amsmath,bm}
\usepackage{mathrsfs}
\usepackage{graphicx,epsfig,braket,amssymb,amsfonts}

\newcommand{\ba}{\begin{eqnarray}}
\newcommand{\ea}{\end{eqnarray}}
\newcommand{\be}{\begin{equation}}
\newcommand{\ee}{\end{equation}}
\newcommand{\bd}{\begin{displaymath}}
\newcommand{\ed}{\end{displaymath}}
\renewcommand{\v}[1]{{\bf #1}}
\newcommand{\bpm}{\begin{pmatrix}}
\newcommand{\epm}{\end{pmatrix}}
\newcommand{\nn}{\nonumber \\}

\begin{document}

\title{Skyrmion Dynamics and Disintegration in Spin-1 Bose-Einstein Condensate}

\author{Xiao-Qiang Xu}
\affiliation{Department of Physics and BK21 Physics Research
Division, Sungkyunkwan University, Suwon 440-746, Korea}
\affiliation{Department of Physics,
Hangzhou Normal University, Hangzhou 310036, China}
\author{Jung Hoon Han}
\email[Electronic address:$~~$]{hanjh@skku.edu}
\affiliation{Department of Physics and BK21 Physics Research
Division, Sungkyunkwan University, Suwon 440-746, Korea}
\affiliation{Asia Pacific Center for Theoretical Physics, POSTECH,
Pohang, Gyeongbuk 790-784, Korea}

\begin{abstract} Dynamics of Skyrmionic spin texture in the spin-1
Bose-Einstein condensate (BEC) is examined by analytical and
numerical means. We show the Skyrmion (coreless vortex) to be
inherently unstable in the sense that the state initially prepared
purely within the anti-ferromagnetic (ferromagnetic) order parameter manifold inevitably evolves into a mixture of both.
The vorticity-dependent drift in the presence of the
trapping potential also contributes to the disintegration of the
initial spin texture manifold. We argue that the notion of a Skyrmion as a
topologically protected entity becomes ill-defined during the dynamical evolution
process.
\end{abstract}


\maketitle

\section{Introduction}

Optical traps enable the possibility to host more than one hyperfine
spin state of cold atoms simultaneously, therefore opening up the popular
study of spinor Bose-Einstein condensate (BEC) since the first
experimental realization~\cite{firstSpinor}. The pioneering theories
of the spin-1 BEC worked out by Ho~\cite{ho}, and independently by
Ohmi and Machida~\cite{machida}, allowed the understanding of the
ground state structures as either  anti-ferromagnetic (AFM) or
ferromagnetic (FM), and elementary excitations. Besides the equilibrium properties, the dynamical
behaviors also received great experimental
attention~\cite{spinDynamics} later on, breeding new research
interests.

The concept of Skyrmion comes from a model for baryons in nuclear
physics~\cite{Skyrme}, and has been regarded as a topological particle of great significance
in condensed matter physics as well. In AFM
spinor BECs we can also predict the existence of the
counterpart which corresponds to a metastable
excitation, while in FM spinor condensate the phrase ``coreless vortex"
is adopted sometimes~\cite{SkyrInBEC-Theo,shin-njp}.
Bogoliubov-de Gennes equations are very useful to study the
stability of such spin textured states in cold atom systems~\cite{virtanen07,machida09}.
Thermal fluctuations, etc,
 can limit the Skyrmion lifetime
which is the key element to make sure we can observe, even manipulate
Skyrmions in BEC system. Previous experiments have already
successfully created the Skyrmion in both AFM
($^{23}$Na)~\cite{ketterle,shin} and FM
($^{87}$Rb)~\cite{SkyrInBEC-Exp-Bigelow} condensates. The decay process into
the ground state was also observed~\cite{shin}. However, few
discussions have been made on the exact temporal dynamics of
Skyrmions before reaching the thermal equilibrium. Here we focus on the pure spin-1 BEC
system at zero temperature, and study the time evolution behaviors
starting from either AFM or FM states supporting the Skyrmionic spin
texture.

The paper is organized as follows. In Sec.~\ref{sec:GS}, we introduce
the concept of Skyrmion (coreless vortex) in the spin-1 condensate for both AFM and FM
cases, and in Sec.~\ref{sec:EoM} we analytically discuss the
possible decay of Skyrmion (coreless vortex) originated from the breakdown of AFM
(FM) phase in the dynamical process. Numerical simulations of the real
time evolution behaviors are made in Sec.~\ref{sec:Numerics} to
show the detailed decay phenomena. Three different initial configurations are
considered, exhibiting different evolution patterns. Finally in
Sec.~\ref{sec:sum}, we summary our results and make corresponding
discussions.

\section{Ground state and Skyrmion}\label{sec:GS}

In the mean-field approximation, the energy functional for the spin-1 condensate without the trapping potential is commonly given
by~\cite{ho,machida}
\ba
E[\Psi] = {\hbar^2 \over 2 m} \int (\bm \nabla \Psi^\dag )\!\cdot\!
(\bm \nabla \Psi )+
 {1\over 2} \int \rho^2 [ c_0 + c_2 (\v S)^2 ],
\ea
where $c_0 = (a_0+2a_2)4\pi\hbar^2/m$ and $c_2 =
(a_2-a_0)4\pi\hbar^2/m$ characterize the interaction strengths, $a_i$
($i=0, 2$) are the $s$-wave scattering length in the two-atom $i$-th scattering
channel, and $m$ is the particle mass. The spin-1 condensate wave
function can be decomposed as
\ba
\Psi (\rho, \theta, \bm \eta ) = \sqrt{\rho} e^{i\theta} \bm
\eta,
\ea
where $\rho$ and $\theta$ are the total density and overall phase,
and $\v S = \bm\eta^\dag \v F \bm\eta$ describes the spin vector with $\bm\eta$ the unit-modulus
spinor field obeying $\bm\eta^\dag \bm\eta = 1$.
The $3\times 3$ angular momentum operator for spin-1 condensate is represented by $\v F = \{ F_x, F_y, F_z \}$.
It is a
common practice to distinguish the two regimes of the ground state
according to $c_2$ being repulsive or attractive. The former corresponds to the AFM case where the spin average becomes zero,
$|\v S|=0$, to minimize the spin-dependent interaction energy, whereas the latter is the FM case and the
spin average is maximized, $|\v S|=1$.
 With the aid of spin
rotation operator $\mathcal{U}(\alpha, \beta, \gamma)= e^{-i F_z
\alpha} e^{-i F_y \beta} e^{-i F_z \gamma}$, where $\alpha$, $\beta$,
and $\gamma$ are the Euler angles, we may express the two fields
as~\cite{ho,machida}
\ba
\bm \eta_1 &=& \mathcal{U} \bpm 0 \\ 1
\\ 0 \epm \!=\! \bpm -{1\over\sqrt{2}} e^{-i\alpha} \sin \beta
\\ \cos\beta \\ {1\over\sqrt{2}} e^{i\alpha} \sin \beta \epm, \nn
\bm \eta_2 &=& \mathcal{U} \bpm 1 \\ 0 \\ 0 \epm \!=\! e^{-i\gamma} \bpm e^{-i\alpha} \cos^2 {\beta \over 2 } \\
{1\over\sqrt{2}} \sin \beta \\ e^{i\alpha} \sin^2 {\beta \over 2 }
\epm . \label{eq:eta1-eta2}
\ea
Throughout the paper we denote $\bm \eta_1$ and $\bm \eta_2$ for AFM
and FM order parameter manifolds, respectively. The two angular
variables $\alpha$ and $\beta$ can be used to construct a unit vector $\v
d= (\sin \beta \cos \alpha, \sin \beta \sin \alpha, \cos \beta)$
which in turn can give rise to a topological spin texture of the
Skyrmion (coreless vortex), where the integer $Q = \int \textrm{d}x \textrm{d}y \, \v d
\cdot (\partial_x \v d \times \partial_y \v d)/(4\pi)$ denotes the
topological charge. Typically, Skyrmions refer to a configuration
where the Euler angle $\alpha$ corresponds to the azimuthal angle
$\phi$ of the plane, and $\beta$ is a function of radial distance $r$
varying from $0$ at the origin, to $\pi$ at infinity or the boundary
of the condensate. Skyrmions can be imprinted in spinor BECs by
a spin rotation method.
In a recent experiment~\cite{shin}, the Skyrmion
created in the AFM $^{23}$Na condensate decays into the ground state
over time, producing what appears to be a half-quantum
vortex-anti-vortex pair~\cite{Justin} within the condensate in the intermediate phase.
Motivated by the dichotomy of the (theoretically) expected
topological stability of the Skyrmion in the AFM BEC and its smooth
decay found in the experiment, it is a timely exercise to carry
out a more critical analysis of the Skyrmion dynamics in the spin-1
condensate.

\section{Equation of motion analysis}\label{sec:EoM}

Some ``pathology" in the spin-1 AFM dynamics was in fact noted early
on~\cite{stoof}. The kinetic energy obtained for the AFM wave
function $\Psi_1$ reads
\ba E_{1} &=& {\hbar^2  \over 2m} \int \rho
\Big[ (\bm \nabla \v d )^2 \!+\! (\bm \nabla \theta )^2 \Big] \!+\!
(\bm \nabla \sqrt{\rho} )^2 \label{eq:H0-AFM}
\ea
while the action part $S_1 =i\int \Psi^\dag_1 \partial_t \Psi_1$
becomes $S_1 = -\int \rho \partial_t \theta$. As it happens, there is
no term in the action responsible for the dynamics of the $\v
d$-vector! By contrast the FM spinor wave function $\Psi_2
= \sqrt{\rho}e^{i\theta}\bm \eta_2$ gives rise to $S_2 = - \int
\rho\partial_t \theta + \int (\cos \beta -1 )
\partial_t \alpha$, where the second term, Berry phase action, is
responsible for the spin dynamics of $\v d$ governed by
Landau-Lifshitz equation of motion. The discrepancy is further
illustrated by the examination of the overlap integral between
adjacently located wave functions $\Psi (\v r )$ and $\Psi (\v r -
\delta \v r)$ for the two cases~\cite{haldane,niu},
\ba
\Psi^\dag_1 (\v r- \delta \v r ) \Psi_1 (\v r) &\simeq& \rho (\v
r)  ,\nn
\Psi^\dag_2 (\v r- \delta \v r ) \Psi_2 (\v r) &\simeq& \rho(\v r
)e^{i \cos \beta (\bm \nabla \alpha \cdot \delta \v r ) } .
\ea
The overlap of the adjacent FM wave functions produces the
Berry phase factor, which is absent for AFM wave function overlap.

A way to cure the pathology of the $\v d$-vector dynamics in the AFM
manifold was suggested by the authors of Ref.~\cite{stoof}, who
considered small fluctuations away from the AFM manifold and
obtained, by integrating out the fluctuations, an effective action
for $\v d$ that is quadratic in time $\sim (\partial \v d /\partial
t)^2$, rather than first-order as expected in superfluid
vortices~\cite{haldane,niu} and magnetic
Skyrmions~\cite{stone,zang}. A similar idea was explored by
Ruostekoski and Anglin, who numerically observed the spontaneous
deformation of the monopole core into an extended defect called the
Alice string, and attributed the phenomenon to the energetic balance
of AFM and non-AFM components in the wave function~\cite{anglin}.

Here we want to offer another perspective of this dynamical problem.
When we define the spinor fields in Eq.~(\ref{eq:eta1-eta2}), we rotated the two
bases $(0,1,0)^T$ and $(1,0,0)^T$, but ignored the third one $(0,0,1)^T$. Euler rotation of this third basis yields
\ba
\bm\eta_3 = \mathcal{U} \bpm 0 \\ 0 \\ 1 \epm = e^{i\gamma} \bpm e^{-i\alpha} \sin^2 {\beta \over 2 } \\
-{1\over\sqrt{2}} \sin \beta \\ e^{i\alpha} \cos^2 {\beta \over 2 }
\epm
\ea
which also corresponds to the FM manifold.
Together the three fields $\bm\eta_j$ ($j=1,2,3$) form a complete, orthogonal set
obeying $\bm\eta^\dag_j \bm\eta_k = \delta_{jk}$.
An arbitrary spinor $\bm\eta$ must therefore be a linear combination of the three basis vectors.
Now we consider the initial AFM or FM state, and let it evolve for a small time step $\Delta t$
to see the composition of the intermediate state, i.e.,
$\Psi(\Delta t) - \Psi(0) \approx - i \Delta t H(0) \Psi(0)$, where
$H = - \hbar^2 \bm\nabla^2 /(2m) + c_0 \Psi^\dag \Psi + c_2 (\Psi^\dag \v F \Psi) \cdot \v F$.

Firstly, for the AFM state we set $\Psi(0) = \bm\eta_1$, and the action of Hamiltonian
on it yields
\ba
H \bm\eta_1 = -\frac{\hbar^2}{2m}\bm\nabla^2 \bm\eta_1 + c_0 \bm\eta_1 = u_j \bm\eta_j,
\ea
with
\ba
u_1 &=& \frac{\hbar^2}{2m} [ (\partial_i \beta)^2 + (\partial_i \alpha)^2 \sin^2 \beta ] + c_0, \nn
u_2 &=& -\frac{\hbar^2}{2m} ( A_1 + A_2 + A_3 - A_4) e^{i\gamma}, \nn
u_3 &=& -\frac{\hbar^2}{2m} ( -A_1 + A_2 + A_3 + A_4) e^{-i\gamma},
\ea
where we have defined $A_1 = (\partial_i \alpha)^2  \sin\beta \cos\beta /\sqrt{2}$,
$A_2 = i (\partial^2_i \alpha)  \sin\beta /\sqrt{2}$, $A_3 = i \sqrt{2} ( \partial_i  \alpha )( \partial_i \beta) \cos\beta$,
and $A_4 = \partial^2_i \beta /\sqrt{2}$. Note that the Einstein convention is used, and $i=x, y$ for two dimensions.
In general, $u_j$ ($j=1,2,3$) are non-zero. As a result,
the initial pure AFM state would evolve into the mixture containing the FM component. The disappearance
of $c_2$ in $u_j$ indicates that $c_2$ plays no role in the beginning of the evolution.

For the FM initial state, $\Psi(0) = \bm\eta_2$, we have
\ba
H \bm\eta_2 = -\frac{\hbar^2}{2m}\bm\nabla^2 \bm\eta_2 + c_0 \bm\eta_2 + c_2 \v d \cdot \v F \bm\eta_2 = v_j \bm\eta_j,
\ea
with
\ba
v_1 &=& -\frac{\hbar^2}{2m} \{ \frac{1}{\sqrt{2}} [2(\partial_i \alpha) (\partial_i \gamma) + i\partial^2_i \gamma]  \sin\beta \nn
  && - \frac{i}{\sqrt{2}} [2(\partial_i \beta)(\partial_i \gamma) + i\partial^2_i \beta] \nn
  && + \frac{1}{\sqrt{2}}(\partial_i \alpha)^2  \sin\beta \cos\beta \} e^{-i\gamma}, \nn
v_2 &=& -\frac{\hbar^2}{2m} (- B_1 - B_2 + B_3 - B_4 - B_5 - B_6) \nn
  && + c_0 + c_2, \nn
v_3 &=& -\frac{\hbar^2}{2m} (-B_1 + B_2 + B_3 + B_4 - B_5 + B_6) e^{-2i\gamma}, \nn
\ea
where we have $B_1 = [(\partial_i \alpha)^2 + (\partial_i \gamma)^2 + i\partial^2_i \alpha] /2$,
$B_2 = [ 2(\partial_i \alpha)(\partial_i \gamma) + i\partial^2_i \gamma] (\cos\beta) /2$,
$B_3 = i(\partial_i \beta)(\partial_i \alpha)  \sin\beta$,
$B_4 = [(\partial_i \beta)^2 + i \partial^2_i \gamma + (\partial_i \gamma)^2]  / 2$,
$B_5 = [2(\partial_i \alpha)(\partial_i \gamma) + i \partial^2_i \alpha]  (\cos\beta) / 2$,
and $B_6 = (\partial_i \alpha)^2 (\cos^2\beta )/ 2$. Immediately, we can draw the conclusion
that the initial pure FM state would also be mixed with AFM component during the time evolution.
Additionally, positive (negative) $c_2$ may enhance (reduce) the effects of $c_0$, which will
be further confirmed later.

The main lesson of the above short-time analysis is that we can readily deduce the source of
AFM-FM mixing to be mainly the inhomogeneity in $\alpha$ and $\beta$,
that is to say, the kinetic energy carried by the inhomogeneous initial condensate wave function. It is the spin
texture in the initial configuration which drives
the mixing. A textured localized object such as a Skyrmion (coreless vortex) in a
pure AFM (FM) condensate inevitably evolves into a mixed phase
due to its intrinsic inhomogeneity.

\section{Description of numerics}\label{sec:Numerics}

Exact time evolution behavior can be obtained by numerically solving the full
set of Gross-Pitaevskii (GP) equations in the presence of a trap
\ba
i {\partial \over \partial t} \Psi \!=\! \left( -{\hbar^2\bm
\nabla^2 \over 2m} + V + c_0 (\Psi^\dag \Psi ) + c_2 (\Psi^\dag \v F
\Psi )\cdot \v F \right)\! \Psi, \nn
\ea
where $V=m\omega^2 \v r^2/2$ ($\v r = x \v e_x + y \v e_y$) is the 2D
harmonic trapping potential. To make all quantities dimensionless we
set the energy, length, and time scales as $\hbar \omega$,
$\sqrt{\hbar/m \omega}$, and $1/\omega$, respectively. As we are
mainly interested in the Skyrmion (coreless vortex) dynamics in the spin-1 condensate,
initial states are chosen to be either $\Psi_{\mathrm{AFM}}(\v r, 0)
= \sqrt{\rho(\v r - \bm\xi_{\rho})}\bm\eta_1(\v r
-\bm\xi_{\mathrm{s}})$ or $\Psi_{\mathrm{FM}}(\v r, 0) =
\sqrt{\rho(\v r - \bm\xi_{\rho})}\bm\eta_2(\v r
-\bm\xi_{\mathrm{s}})$ which represent the Skyrmion (coreless vortex) in pure
AFM (FM) state. For general consideration
we introduce the displacement of the density peak from the trap center
by $\bm\xi_{\rho}$, while
$\bm\xi_{\mathrm{s}}$ is the displacement of Skyrmionic spin texture.
We assume both $\bm\xi_{\rho}$ and
$\bm\xi_{\mathrm{s}}$ are small quantities (compared to the condensate size) in the $x$ direction, i.e., $\bm\xi_{\rho}
= \xi_{\rho} \v e_x$, and $\bm\xi_{\mathrm{s}} = \xi_{\mathrm{s}} \v
e_x$. The density distribution takes the Gaussian profile $\rho(\v r
- \bm\xi_{\rho})=e^{-(\v r-\bm\xi_{\rho})^2}/\sqrt{2\pi}$ which is
the single particle ground state in the harmonic trapping potential.
Our choice of the density profile is not the exact many-body ground (or metastable) state of the given Hamiltonian.
For the sake of evolution, derivation from such exact state is desired, therefore we choose the current form
for simplicity.
For the AFM spinor fields we have
\ba
\bm\eta_1 = \bpm -{1\over\sqrt{2}} e^{-i\phi(\v r -
\bm\xi_{\mathrm{s}})} \sin \beta(\v r - \bm\xi_{\mathrm{s}})
\\ \cos\beta(\v r - \bm\xi_{\mathrm{s}})
\\ {1\over\sqrt{2}} e^{i\phi(\v r - \bm\xi_{\mathrm{s}})} \sin \beta(\v r - \bm\xi_{\mathrm{s}}) \epm,
\ea
where $\phi(\v r- \bm\xi_{\mathrm{s}})$ is the
azimuthal angle, and $\cos \beta(\v r - \bm\xi_{\mathrm{s}}) = [(\v r
- \bm\xi_{\mathrm{s}})^2-1]/[(\v r - \bm\xi_{\mathrm{s}})^2+1]$,
$\sin \beta(\v r - \bm\xi_{\mathrm{s}}) = 2|\v r -
\bm\xi_{\mathrm{s}}| /[(\v r - \bm\xi_{\mathrm{s}})^2+1]$ is used.
For the FM case, in view of the recent
experiment~\cite{SkyrInBEC-Exp-Bigelow}, we may adopt
\ba
\bm\eta_2 = \bpm \cos^2 \frac{\beta(\v r- \bm\xi_{\mathrm{s}})}{2} \\
    \frac{1}{\sqrt{2}} e^{i\phi(\v r - \bm\xi_{\mathrm{s}})}\sin \beta(\v r- \bm\xi_{\mathrm{s}})  \\
     e^{2i\phi(\v r - \bm\xi_{\mathrm{s}})} \sin^2 \frac{\beta(\v r- \bm\xi_{\mathrm{s}})}{2} \epm,
     \label{eq:GP-FM}
\ea
where $\beta(\v r - \bm\xi_{\mathrm{s}}) = \pi(1-e^{-|\v r-
\bm\xi_{\mathrm{s}}|})$ varies from $0$ to $\pi$. Note that
similar results are always obtained regardless of the choice of the exact radial dependence of $\beta$.
We may define $\Psi = (\psi_{+1}, \psi_0, \psi_{-1})^T$,
where $\psi_i$ ($i=+1, 0, -1$) denote the wavefunctions in the corresponding hyperfine states.
Initially, the
winding numbers of $\psi_{+1}$, $\psi_0$, and $\psi_{-1}$ for AFM and
FM cases are $(-1,0,1)$ and $(0,1,2)$, respectively. In the following
calculations, the real time evolution method is used, and we set $c_0
= 50.0$. Considering the possible experimental setup, we discuss the
following three kinds of situations with regard to different choices of
$\xi_\rho$ and $\xi_{\mathrm{s}}$.

\subsection{$\xi_{\rho}=\xi_{\mathrm{s}}=0$}

\begin{figure}[tbph]
\centering
\includegraphics[width=80mm]{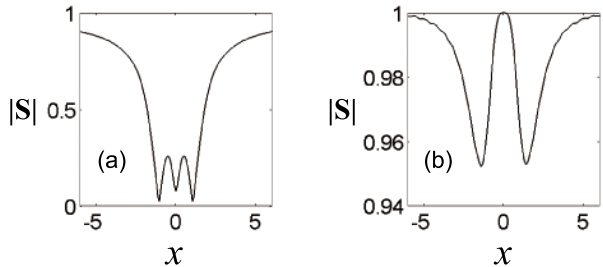}
\caption{Radial dependence of $|\v S (x,y=0, t)|$ for the initial
Skyrmionic configuration defined on (a) AFM and (b) FM manifolds taken
at time $t=2.0$ with $c_2=0.0$. Other values of $c_2$ from -20 to 20 were tried, supporting similar results.}\label{fig:A}
\end{figure}

In this case, both the density and the Skyrmionic texture are
centered, enabling the rotational symmetry to be preserved during the whole evolution process.
As discussed in Sec.~\ref{sec:EoM}, either a pure
AFM or a pure FM initial state will evolve into a mixture of
both components. Indeed, as shown in Fig.~\ref{fig:A}(a), non-zero
$|\v S (\v r, t)|$ ($t>0$) indicates the appearance of the FM
feature out of the AFM initial state, while in Fig.~\ref{fig:A}(b), the
deviation of $|\v S (\v r, t)|$ from 1 shows the loss of FM
character although the initial state was fully ferromagnetic.

Naturally it becomes problematic to give the precise meaning to the
Skyrmion now that the very manifold (either AFM or FM) on which it
was defined gets disintegrated over time. It might be possible,
through decomposition of the condensate wave function in the basis space
spanned by $\bm\eta_j$ $(j=1,2,3)$ as introduced in the previous section,
to deduce the $\v d (\v r, t)$-vector and the Skyrmion density
$\v d \cdot (\partial_x \v d \times \partial_y \v d)/4\pi$
formally at any given time $t$ for any wave function $\Psi(\v r,
t)$. This is a subtle and important issue which we will treat in a later publication.  In the mixed AFM and FM state,
the winding numbers of $\psi_i$ ($i=+1, 0, -1$) are still preserved during
the evolution, and the vortex (anti-vortex) cores always coincide
with each other. Since some Skyrmionic features continue to be maintained over time,
we may name this structure in the mixed state as a ``pseduo-Skyrmion" for the
convenience of description.

\subsection{$\xi_{\rho}=0$ and $\xi_{\mathrm{s}}>0$}

In the previous subsection the initial Skyrmionic configuration obeyed
the rotational symmetry. The subsequent disintegration of the AFM or FM
component was largely due to the inherent dynamics of the
condensate, unrelated to effects arising from the trap potential.
Now we consider the case when the Skyrmion (coreless vortex) is initially displaced
from the trap center.
In the spin rotation method as adopted in Ref.~\cite{shin}, by
tuning the position of the zero-field center of the 3D quadrupole magnetic field,
we may manipulate the position of the Skyrmion (coreless vortex). Therefore, our setup
is experimentally feasible. As a comparison, for a single component condensate it is well
known that a displaced vortex would precess around the trap core
driven by a radial buoyant force and a gyroscopic Magnus force together~\cite{Precession}.

\begin{figure}[tbph]
\centering
\includegraphics[width=85mm]{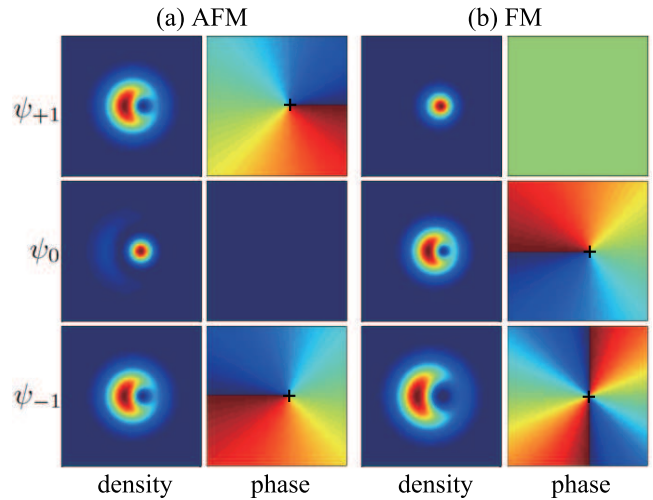}
\caption{(color online) Initial density and phase profiles for a displaced Skyrmion within the
AFM manifold (a), and those of a displaced coreless vortex within the
FM manifold (b). The first, second, and third rows correspond to
$\psi_{+1}$, $\psi_0$, and $\psi_{-1}$, respectively. Initial displacements are both at
$\xi_{\mathrm{s}}=0.5$. The plus signs (+)
mark the positions of vortex (anti-vortex) cores. }\label{fig:initialB}
\end{figure}

\begin{figure}[tbph]
\centering
\includegraphics[width=85mm]{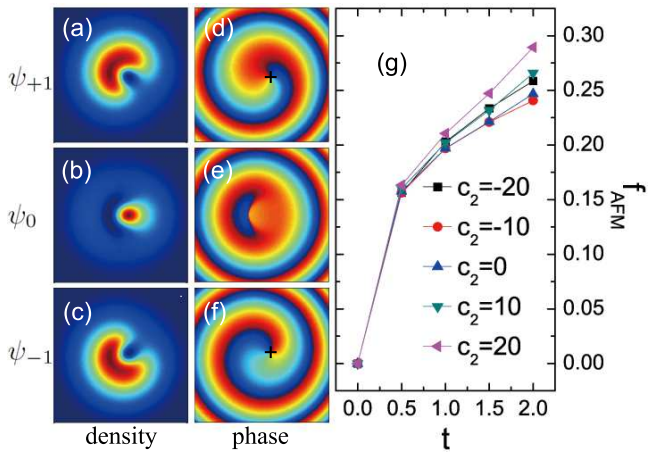}
\caption{(color online) Real time evolution starting from a displaced Skyrmion within the
AFM initial manifold as shown in Fig.~\ref{fig:initialB}(a) for $c_2=0.0$. (a)-(c) Snapshots of densities of $\psi_{+1}$,
$\psi_0$, and $\psi_{-1}$, respectively, at $t=2.0$. (d)-(f) are the
corresponding phases.  The plus signs (+)
mark the positions of vortex (anti-vortex) cores. (g) Time dependence of
$f_{\mathrm{AFM}}(t)$ for different values of $c_2$. }\label{fig:B-AFM}
\end{figure}

For our spin-1 AFM initial state as shown in Fig.~\ref{fig:initialB}(a), the anti-vortex of $\psi_{+1}$ and
vortex of $\psi_{-1}$ are located at the same position
$\bm\xi_{\mathrm{s}}$. However, they feel the opposite forces, and
move in opposite directions as shown in Fig.~\ref{fig:B-AFM}(a)-(f).
Note that similar behavior in the displaced AFM monopole dynamics of
spin-1 condensate was discussed in Ref.~\cite{Martikainen}. The
splitting of the vortex and the anti-vortex cores implies
that description of the condensate wave function based purely on AFM
manifold is no longer tenable. To quantify the breakdown of the AFM
phase, we define
\ba
f_{\mathrm{AFM}}(t) = \frac{\int \textrm{d}x \textrm{d}y \,
||\psi_{+1}(\v r, t)|^2 - |\psi_{-1}(\v r, t)|^2|} {\int \textrm{d}x
\textrm{d}y \, ||\psi_{+1}(\v r, t)|^2 + |\psi_{-1}(\v r, t)|^2|}
\ea
as a measure of the loss of AFM component. Pure AFM condensate would
obey $|\psi_{+1}(\v r, t)|^2 - |\psi_{-1}(\v r, t)|^2|=0$. In
Fig.~\ref{fig:B-AFM}(g) the obvious trend of increase of
$f_{\mathrm{AFM}}(t)$ from zero confirms our picture that vortex-anti-vortex
core separation implies the loss of AFM condensate.
As $t \rightarrow 0$, the effects of $c_2$ caused by its value and sign
are reduced as clear from the figure, confirming the analysis in Sec.~\ref{sec:EoM}.

The simulation results suggest another way for the initial Skyrmion
structure to disintegrate. The inherent dynamics of the vortex in the
superfluid follows that of electrons in quantized magnetic
field~\cite{haldane,niu}. The trapping potential provides the
confining force, the analogue of electric field, for the vortex so
that the drift motion orthogonal to the confining force occurs. The
charge of the vortex is opposite to that of the anti-vortex, so the
drifts occur in opposite directions. Unless there exists an enormous
force binding the vortex and the anti-vortex cores together, the
drift mechanism will inevitably separate them in opposite directions
and results in disintegration of the Skyrmion.

\begin{figure}[tbph]
\centering
\includegraphics[width=85mm]{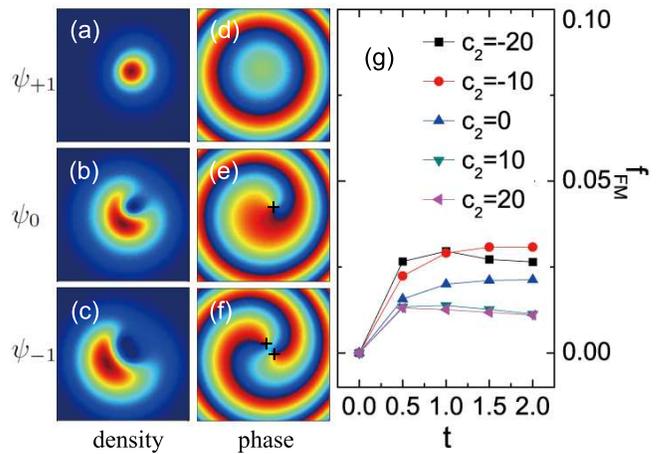}
\caption{(color online) Real time evolution starting from a displaced coreless vortex within the
FM initial manifold as shown in Fig.~\ref{fig:initialB}(b) for $c_2=0.0$. (a)-(c) Snapshots of densities of $\psi_{+1}$, $\psi_0$,
$\psi_{-1}$, respectively, at $t=2.0$. (d)-(f) are the corresponding phases.
The plus signs (+)
mark the positions of vortex cores. (g)
Time dependence of $f_{\mathrm{FM}}(t)$ for different values of
$c_2$. }\label{fig:B-FM}
\end{figure}

As for the FM initial state in Eq.~(\ref{eq:GP-FM}), both vortices in
$\psi_0$ and $\psi_{-1}$ have the same sign of the vorticity as seen from
Fig.~\ref{fig:initialB}(b). Therefore,
their cores move in the same direction  when initially displaced from
the trap center, as shown in Fig.~\ref{fig:B-FM}(a)-(f). During the
evolution we observe that the $l = 2$ vortex in $\psi_{-1}$ is
unstable and decays into two $l = 1$ vortices, as shown in
Fig.~\ref{fig:B-FM}(c) and \ref{fig:B-FM}(f)~\cite{virtanen07,machida09}. To characterize the
decay of the FM feature we define~\cite{comment}
\ba
f_{\mathrm{FM}}(t) \! = \! \! \frac{\int \textrm{d}x \textrm{d}y
(|\psi_{+1}(\v r, t)| + |\psi_{-1}(\v r, t)| -\sqrt{\rho})^2} {\int
\textrm{d}x \textrm{d}y (|\psi_{+1}(\v r, t)| + |\psi_{-1}(\v r, t)|
+\sqrt{\rho})^2}.
\ea
The definition is motivated by the observation that, in perfect FM
state, $|\psi_{+1}(\v r, t)| + |\psi_{-1}(\v r, t)| =\sqrt{\rho}$ is
guaranteed. From Fig.~\ref{fig:B-FM}(g) we can observe the obvious
trend of increase of $f_\mathrm{FM}(t)$ from zero. Another conclusion
from the figure is that
the positive $c_2$ may enhance the stiffness of the initial FM manifold, while
the negative $c_2$, on the contrary, would favor faster speed of the
disintegration from the FM state.

\subsection{$\xi_{\rho}=\xi_{\mathrm{s}}>0$}

\begin{figure}[tbph]
\centering
\includegraphics[width=70mm]{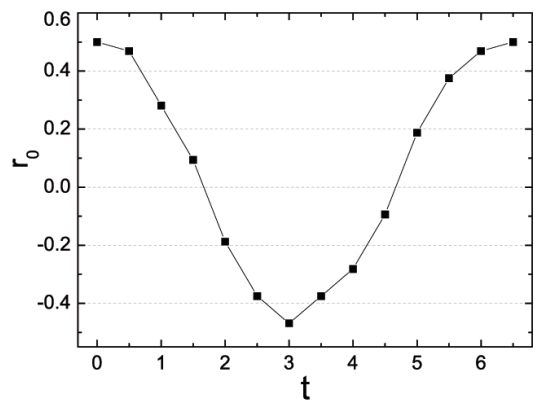}
\caption{Time dependence of the pseudo-Skyrmion displacement from the trap
center for both AFM and FM initial states. The difference between the two is too
small to distinguish here. Initial displacements are
$\xi_{\rho}=\xi_{\mathrm{s}}=0.5$, and $c_2=0.0$. Other values of
$c_2$ from -20 to 20 lead to similar quasi-periodic
behaviors.}\label{fig:C}
\end{figure}

Now we consider the case when the center of the density is displaced together with that of the
Skyrmionic texture. This situation is also realizable in experiments. Firstly, we
create a Skyrmion (coreless vortex) centered in the harmonic trap following the standard
procedure. After that we suddenly
shift the trapping potential, therefore both the density and the Skyrmion (coreless vortex) core would
be displaced with respect to the new trap center. During the time
evolution the whole condensate moves like a pendulum about the trap
center, accompanied by the disintegration of initial AFM or FM states. Contrary to the
splitting of vortex-anti-vortex pair or the breakdown of $l=2$ vortex seen in the previous subsection,
now the pseudo-Skyrmion structure is always protected throughout the evolution for both
two initial configurations. We show in Fig.~\ref{fig:C} the
time dependence of the displacement of the pseudo-Skyrmion from the trap
center, where the periodic behavior with only one period of oscillation is
presented. We may approximately describe the periodic motion of the
pseudo-Skyrmion as the semi-classical Newtonian equation of
motion~\cite{stoof}.
  We conclude that here the dynamics of the density
$\dot{\rho}$ dominates over that of the spin texture and effectively
acts as a binding force tying the vortex and anti-vortex cores
together. Due to the nonlinear interactions, deviation from the perfect
periodicity is observed.

\section{Summary and Discussion}\label{sec:sum}

Motivated by the recent experiment of Skyrmion creation and its
evolution in the AFM spin-1 condensate of
$^{23}$Na~\cite{shin}, we have carried out a detailed study of the
Skyrmion (coreless vortex) dynamics based on the GP equation analysis. Contrary to the
standard notion of a Skyrmion as a metastable topological object
in, e.g. non-linear $\sigma$-model~\cite{rajaraman}, it is
\textit{inherently unstable in the BEC condensate} due to the
\textit{dynamical mixing} of the AFM and FM components during the time evolution. The
instability has a different character from that expected in thermal
or quantum tunneling of the topological object out of the metastable
minimum (``fate of the false vacuum")~\cite{coleman}.
Interaction effects are also examined during the early stage of time evolution,
exhibiting the explicit role of $c_2$ in the FM case and the independence of $c_2$
in the AFM counterpart.


Numerically, three feasible initial configurations have been investigated for their dynamical evolution behaviors.
(i) For the Skyrmion (coreless vortex) initially prepared in a pure
AFM (FM) manifold and located at the trap center, the
mixing with the FM (AFM) component is observed.
The rotational symmetric structure is preserved over time. (ii) When we shift the Skyrmion
while keeping the density peaked at the center, splitting of the
vortex and anti-vortex centers is observed for AFM initial state.
Note that the AFM manifold breaks down at $t>0$. For the FM initial coreless vortex state, the $l=2$ vortex breaks up into two
$l=1$ vortices, while rotating around the trap center in the same
direction. This also contributes to the breakdown of the FM manifold. (iii) If both the density peak and the
Skyrmion (coreless vortex) are
displaced by the same amount, the whole condensate including the
pseudo-Skyrmion oscillates through the trap center in analogy to
the pendulum motion. The dynamics is governed by that of the density
oscillation as a whole, not by those of individual vortices.

Given that our results rely on the energy-conserving simulation,
they could not be applied directly to explain the transition into the ground state in Ref.~\cite{shin}.
However, before the system reaches its thermal equilibrium, short time behaviors as
discussed in our paper may be observable in experiments.

\acknowledgments H. J. H. is supported by NRF grant (No.
2010-0008529, 2011-0015631). We acknowledge useful
communication with Y. Shin on the Skyrmion decay and his experimental
input.


\begin{thebibliography}{99}

\bibitem{firstSpinor} D. M. Stamper-Kurn, M. R. Andrews, A. P. Chikkatur,
S. Inouye, H.-J. Miesner, J. Stenger, and W. Ketterle,
Phys. Rev. Lett. \textbf{80}, 2027 (1998).

\bibitem{ho} T.-L. Ho, Phys. Rev. Lett. \textbf{81}, 742 (1998).

\bibitem{machida} T. Ohmi and K. Machida, J. Phys. Soc. Jpn. \textbf{67}, 1822
(1998).

\bibitem{spinDynamics} M.-S. Chang, Q. Qin, W. Zhang, L. You, and M. S. Chapman, Nat. Phys. \textbf{1}, 111 (2005).

\bibitem{Skyrme} T. H. R. Skyrme, Proc. R. Soc. London, Ser. A \textbf{260}, 127 (1961); Nucl. Phys. \textbf{31}, 556 (1962).

\bibitem{SkyrInBEC-Theo} U. A. Khawaja and H. T. C. Stoof, Nature (London) \textbf{411}, 918 (2001);
Phys. Rev. A \textbf{64}, 043612 (2001); H. Zhai, W. Q. Chen, Z. Xu, and L. Chang, Phys. Rev. A \textbf{68}, 043602 (2003).


\bibitem{shin-njp} J. Choi, W. J. Kwon, M. Lee, H. Jeong, K, An, adn Y. Shin, New J. Phys. \textbf{14}, 053013 (2012).

\bibitem{virtanen07} V. Pietil\"{a}, M. M\"{o}tt\"{o}nen, and S. M. M. Virtanen, Phys. Rev. A \textbf{76}, 023610 (2007).

\bibitem{machida09} M. Takahashi, V. Pietil\"{a}, M. M\"{o}tt\"{o}nen, T. Mizushima, and K. Machida, Phys. Rev. A \textbf{79}, 023618 (2009).


\bibitem{ketterle} A. E. Leanhardt, Y. Shin, D. Kielpinski, D. E.
Pritchard, and W. Ketterle, Phys. Rev. Lett. \textbf{90}, 140403
(2003).

\bibitem{shin} J. Y. Choi, W. J. Kwon, and Y. I. Shin, Phys. Rev. Lett. \textbf{108}, 035301 (2012).


\bibitem{SkyrInBEC-Exp-Bigelow} L. S. Leslie, A. Hansen, K. C. Wright, B. M. Deutsch,
and N. P. Bigelow, Phys. Rev. Lett. \textbf{103}, 250401 (2009).


%

\bibitem{Justin} J. Lovegrove, M. O. Borgh, and J. Ruostekoski, Phys. Rev. A, \textbf{86}, 013613 (2012).


\bibitem{stoof} H. T. C. Stoof, E. Vliegen, and U. Al Khawaja, Phys.
Rev. Lett. \textbf{87}, 120407 (2001).

\bibitem{haldane} F. D. M. Haldane and Y.-S. Wu,
Phys. Rev. Lett. \textbf{55}, 2887 (1985).

\bibitem{niu} Q. Niu, P. Ao, and D. J. Thouless, Phys. Rev. Lett. \textbf{72}, 1706 (1994).

\bibitem{stone} M. Stone, Phys. Rev. B \textbf{53}, 16573 (1996).

\bibitem{zang} J. Zang, M. Mostovoy, J. H. Han, and N. Nagaosa,
Phys. Rev. Lett. \textbf{107}, 136804 (2011).

\bibitem{anglin} J. Ruostekoski and J. R. Anglin, Phys. Rev. Lett.
\textbf{91}, 190402 (2003).


\bibitem{Precession} D. V. Freilich, D. M. Bianchi, A. M. Kaufman, T. K. Langin, and
D. S. Hall, Science \textbf{329}, 1182 (2010).

\bibitem{Martikainen} J.-P. Martikainen, A. Collin, and K.-A. Suominen, Phys. Rev. Lett. \textbf{88}, 090404 (2002).

\bibitem{comment} The choice of $f_\mathrm{AFM}(t)$ and
$f_\mathrm{FM}(t)$ is not unique. Other definitions measuring AFM or
FM fractions can be given, presumably with similar results. The
arbitrariness is obviously related to the lack of well-defined
procedure to extract AFM and FM components out of the spin-1 spinor
wave function.

\bibitem{rajaraman} R. Rajaraman, \textit{Solitons and Instasntons}
(North-Holland, Amsterdam, 1987), Chap. 3.

\bibitem{coleman} S. Coleman, \textit{Aspects of Symmetry}
(Cambridge University Press, 1988), Chap. 7.


\end{thebibliography}
\end{document}